\documentclass[conference,10pt, final, twocolumn]{IEEEtran}
\usepackage{ifpdf}
 \ifpdf
 \else
 \fi

\ifCLASSINFOpdf
   \usepackage[pdftex]{graphicx}
\else
   \usepackage[dvips]{graphicx}
\fi

\usepackage[cmex10]{amsmath}
\usepackage{cite}
\usepackage[tight,footnotesize]{subfigure}
\usepackage{define-notations,amssymb}
\usepackage{color,algorithmic,algorithm,cite}

\renewcommand\footnotemark{}

\begin{document}
\title{Novel Subcarrier-pair based Opportunistic DF Protocol for Cooperative Downlink OFDMA}

\author{Tao Wang$^1$,~{\it Senior Member, IEEE}, Yong Fang$^1$,~{\it Senior Member, IEEE},
        and Luc Vandendorpe$^2$,~{\it Fellow, IEEE} \\
        $^1$School of Communication \& Information Engineering, Shanghai University, 200071 Shanghai, P. R. China \\ $^2$ICTEAM Institute, Universit\'e catholique de Louvain, 1348 Louvain-la-Neuve, Belgium \\
        t.wang@ieee.org, yfang@staff.shu.edu.cn, luc.vandendorpe@uclouvain.be
}



\maketitle

\begin{abstract}
A novel subcarrier-pair based opportunistic DF protocol
is proposed for cooperative downlink OFDMA transmission aided
by a decode-and-forward (DF) relay.
Specifically, user message bits are transmitted in two consecutive equal-duration time slots.
A subcarrier in the first slot can be paired with a subcarrier in the second slot
for the DF relay-aided transmission to a user.
In particular, the source and the relay can transmit
simultaneously to implement beamforming at the subcarrier in the second slot
for the relay-aided transmission.
Each unpaired subcarrier in either the first or second slot is used by the source
for direct transmission to a user without the relay's assistance.
The sum rate maximized resource allocation (RA) problem
is addressed for this protocol under a total power constraint.
It is shown that the novel protocol leads to a maximum sum rate
greater than or equal to that for a benchmark one,
which does not allow the source to implement beamforming
at the subcarrier in the second slot for the relay-aided transmission.
Then, a polynomial-complexity RA algorithm is developed
to find an (at least approximately) optimum resource allocation 
(i.e., source/relay power, subcarrier pairing and assignment to users)
for either the proposed or benchmark protocol.
Numerical experiments illustrate that the novel protocol can
lead to a much greater sum rate than the benchmark one.\footnote{
Research supported by The Program for Professor of
Special Appointment (Eastern Scholar) at Shanghai Institutions of Higher Learning.
It is also supported by the IAP project BESTCOM,
the ARC SCOOP, and the NEWCOM\# and NSF China \#61271213.}
\end{abstract}


%
\IEEEpeerreviewmaketitle

\section{Introduction}

The incorporation of subcarrier-pair based decode-and-forward (DF)
relaying into orthogonal frequency division modulation (OFDM) or
multiple-access (OFDMA) transmission
and associated resource allocation (RA) were studied in
\cite{WangYing07,Li08,Haj11,Vandendorpe08-1}
when the source-to-destination (S-D) link exists.
In \cite{WangYing07,Li08,Haj11}, an ``always-relaying" DF protocol was used,
i.e., a subcarrier in the first time slot is always
paired with a subcarrier in the second slot for the relay-aided transmission.
To better exploit the frequency-selective fading,
we have proposed an opportunistic DF relaying protocol (sometimes termed
as selection relaying) in \cite{Vandendorpe08-1,ZhiWen12,ZhiWen13},
i.e, a subcarrier in the first time slot can either be
paired with a subcarrier in the second slot for the relay-aided transmission,
or used for the S-D direct transmission without the relay's assistance.
It is very important to note that when some subcarriers
in the first slot are used for the direct transmission,
some subcarriers in the second slot will not be used,
which wastes spectrum resource.

To address the above issue, we have proposed
an improved DF protocol in\cite{Vandendorpe08-2}.
This protocol is the same as those considered in
\cite{Vandendorpe08-1} except that the source can also make direct S-D transmission
at every unpaired subcarrier in the second slot.
This protocol and its RA were later intensively studied, e.g.,
in\cite{Vandendorpe08-2,Vandendorpe09-2,WangTSP11,WangJSAC11,Hsu11,Boost11}.
Note that {\it the improved protocol does not really
improve the way how DF relaying is implemented over a subcarrier pair,
but rather let the source utilize the unpaired subcarriers in the second slot
for direct transmission to avoid the waste of spectrum resource}.
In \cite{Hsu11,Boost11,Wang13CL}, the subcarrier pairing and power allocation
are jointly optimized for point-to-point OFDM transmission.
As for OFDMA systems, RA problems considering the joint optimization
of power allocation and subcarrier assignment to users are addressed in \cite{Vandendorpe08-2,Vandendorpe09-2,WangTSP11,WangJSAC11}.
In these works, a priori and CSI-independent subcarrier pairing
is considered, i.e., a subcarrier
in the first slot is always paired with the same subcarrier in the second slot
if the relay-aided mode is used.
It is a complicated RA problem to jointly optimize subcarrier pairing,
power allocation and subcarrier assignment to users.

In this paper, we consider downlink OFDMA transmission
from a source to multiple users aided by a DF relay.
Compared with the existing works, this paper makes the following contributions.
First, a novel subcarrier-pair based opportunistic DF relaying protocol is proposed.
A benchmark protocol using the improved protocol 
as in \cite{Vandendorpe08-2} is also considered.
However, the proposed protocol uses further improved relay-aided transmission,
which allows the source and relay to transmit simultaneously
to implement beamforming at the subcarrier in the second slot.
{\it Note that the proposed protocol truly improves the implementation of
DF relaying over a subcarrier pair with transmit beamforming,
which is not the case for the benchmark protocol.}
Second, the sum rate maximized RA problem
is addressed for both the novel and benchmark protocols,
under a total power constraint for the whole system.
It is shown that the novel protocol leads to a maximum sum rate greater than
or equal to that for the benchmark one.
An RA algorithm is developed for each protocol
to find the globally optimum source/relay power allocation
and subcarrier pairing to maximize the sum rate of all users.

The rest of this paper is organized as follows.
In the next section, the system and transmission protocols are described.
In Section \ref{sec:insight}, we will focus on computing the maximum rate
and optimum power allocation for a subcarrier pair
using the relay-aided transmission for both protocols. Using these results,
an RA algorithm will be developed in Section \ref{sec:RA},
and numerical experiments are shown in Section \ref{sec:numexp}.
Finally, some conclusions are drawn.

Notations: A letter in bold, e.g. $\bf x$, represents a set.
$\Rate{x} = \frac{1}{2}\log_2(1 + x)$ and $[x]^+ = \max\{x,0\}$.

\section{Protocols and WSR maximization problem}

\subsection{The transmission system and protocols}

Consider the downlink OFDMA transmission from a source to $U$ users collected
in the set $\Uset = \{u|u=1,\cdots,U\}$ aided by a DF relay.
The source and the relay can simultaneously emit
OFDM symbols using $K$ subcarriers and with sufficiently long cyclic prefix
to eliminate inter-symbol interference.
User message bits are transmitted in two consecutive
equal-duration time slots, during which all channels are assumed to keep unchanged.
During the first slot, only the source broadcasts $N$ OFDM symbols.
Both the relay and all users receive these symbols.
After proper processing explained later, the source and relay simultaneously
broadcast $N$ OFDM symbols, and the users receive them during the second slot.
Due to the OFDMA, each subcarrier is dedicated to transmitting
a single user's message exclusively.
A subcarrier in the first slot can be paired with a subcarrier in the second slot
for the relay-aided mode transmission to a user.
Each unpaired subcarrier in either the first or second slot is used by the source
for the direct mode transmission to a user.

To simplify description, we use subcarriers $k$ and $l$ to denote
the $k$th and $l$th subcarriers used during the first and second slots, respectively ($k,l=1,\cdots,K$).
We define the source transmission powers for subcarrier $k$
in the first slot and subcarrier $l$ in the second slot as $\Pskone$ and $\Psltwo$, respectively.
The relay transmission power for subcarrier $l$ is $\Prl$.
The complex amplitude gains at subcarrier $k$
for the source-to-relay, source-to-$u$ and relay-to-$u$ channels
are $\hsrk$, $\hsuk$ and $\hruk$, respectively.
The two transmission modes for the novel protocol are elaborated as follows:

\medskip
\subsubsection{The relay-aided transmission mode}

\begin{figure}
  \centering
  \subfigure[]
  {\includegraphics[width=1.7in]{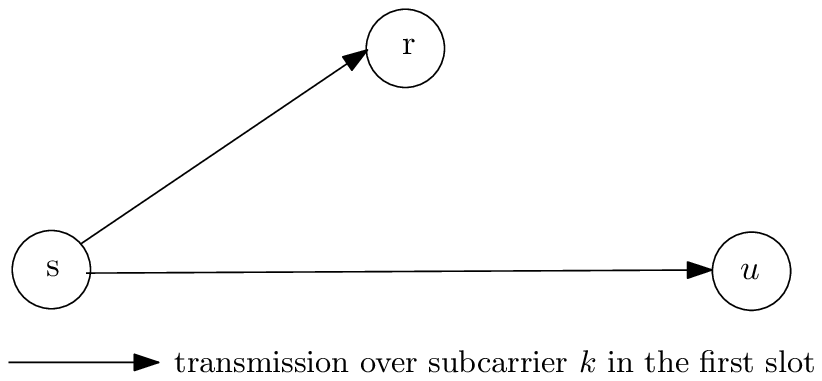}}
  \subfigure[]
  {\includegraphics[width=1.7in]{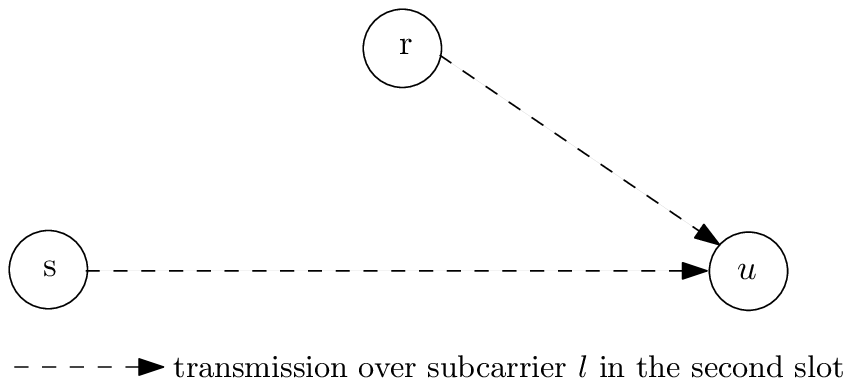}}
  \caption{The relay-aided transmission mode over the subcarrier pair $(k,l)$ to user $u$.}\label{fig:relay-mode}
\end{figure}

Suppose subcarrier $k$ is paired with subcarrier $l$
for the relay-aided mode transmission to user $u$.
A block of message bits are first encoded into a code word of complex symbols
$\{\Code(n)|n=1,\cdots,N\}$ with $E(|\Code(n)|^2)=1$, $\forall\;n$.
In the first slot, the source broadcasts the codeword over subcarrier $k$
as illustrated in Figure \ref{fig:relay-mode}.a.
At the relay and user $u$, the $n$th baseband signals received through subcarrier $k$ are
\begin{align}
\Yrk(n) = \sqrt{\Pskone} \hsrk\Code(n) + \Nrk(n), n = 1,\cdots,N,
\end{align}
and
\begin{align}
\Yukone(n) = \sqrt{\Pskone}\hsuk\Code(n) + \Nukone(n), n = 1,\cdots,N,
\end{align}
respectively, where $\Nrk(n)$ and $\Nukone(n)$ are both additive white Gaussian noise (AWGN)
with power $\sigma^2$. The signal-to-noise ratio (SNR) at the relay is
$\Pskone\Gsrk$ where $\Gsrk=\frac{|\hsrk|^2}{\sigma^2}$.
At the end of the first time slot, the relay decodes the message bits
from $\{\Yrk(n)|n = 1,\cdots,N\}$ and
then reencodes those bits into the same codeword as the source did.

In the second time slot, the source and relay broadcast the codewords
$\{\Code(n)e^{-j\angle\hsul}|\forall\;n\}$ and $\{\Code(n)e^{-j\angle\hrul}|\forall\;n\}$
through subcarrier $l$, respectively, where $\angle\hsul$ and $\angle\hrul$ represent
the phase of $\hsul$ and $\hrul$, respectively.
This means that the source and relay implement transmit beamforming
to emit the codeword through subcarrier $l$ as illustrated in Figure \ref{fig:relay-mode}.b.
Note that the source and relay need to know
the phase of $\hsul$ and $\hrul$, respectively.
At user $u$, the $n$th baseband signal received through subcarrier $l$ is
\begin{align}
\Yultwo(n) = \big(\sqrt{\Psltwo}|\hsul|+\sqrt{\Prl}|\hrul|\big)\Code(n) + \Nultwo(n),
\end{align}
where $\Nultwo(n)$ is the AWGN with power $\sigma^2$.

Finally, user $u$ decodes the message bits from all signals received during the two slots.
These signals can be grouped into $N$ vectors, the $n$th of which is
\begin{align}
\Yn &= \left[\begin{array}{c}
                \Yukone(n) \\
                \Yultwo(n) \end{array}\right]      \label{eq:eqv-channel}  \\
    &= \left[\begin{array}{c}
           \sqrt{\Pskone}\hsuk  \\
           \sqrt{\Psltwo}|\hsul|+\sqrt{\Prl}|\hrul|  \end{array}\right] \Code(n) + \Nn,   \nonumber
\end{align}
where $\Nn = [\Nukone(n),\Nultwo(n)]^T$.
Note that the transmission of the codeword in effect
makes $N$ uses of a discrete memoryless single-input-two-output
channel specified by \eqref{eq:eqv-channel}, with the $n$th input and output being $\Code(n)$ and $\Yn$, respectively.
To achieve the maximum reliable transmission rate,
maximum ratio combining should be used \cite{Fund-WCOM},
It can readily be derived that the SNR for after this combining is
\begin{align}
&\SNRklu(\Pskone,\Psltwo,\Prl) = \Gsuk\Pskone +   \nonumber\\
&\hspace{2cm}         \big(\sqrt{\Gsul\Psltwo}+\sqrt{\Grul\Prl}\big)^2,
\end{align}
where $\Gsuk=\frac{|\hsuk|^2}{\sigma^2}$ and $\Grul=\frac{|\hrul|^2}{\sigma^2}$.
To ensure both the relay and user $u$ can reliably decode the message bits,
the maximum number of message bits that can be transmitted is $2N\Rate{\Gsrk\Pskone}$
and $2N\Rate{\SNRklu(\Pskone,\Psltwo,\Prl)}$, respectively.
This means that the maximum transmission rate over the subcarrier pair $(k,l)$
in the relay-aided mode to user $u$ is equal to
$\Rate{\min\{\Gsrk\Pskone, \SNRklu(\Pskone,\Psltwo,\Prl)\}}$ bits/OFDM-symbol (bpos).

\subsubsection{The direct transmission mode}

Suppose subcarrier $k$ (respectively, subcarrier $l$) is unpaired with any subcarrier
in the second (respectively, first) slot, and is used for direct mode transmission to user $u$.
The source first encodes message bits into a codeword of $N$ symbols,
which are then broadcast through subcarrier $k$ (respectively, subcarrier $l$.
In such a case, the relay keeps silent at subcarrier $l$ in the second slot, i.e. $\Prl=0$.).
User $u$ decodes the message bits from the signals received through subcarrier $k$
(respectively, subcarrier $l$).
The maximum rate through subcarrier $k$ (respectively, subcarrier $l$) in the direct transmission mode
is $\Rate{\Pskone\Gsuk}$ (respectively, $\Rate{\Psltwo\Gsul}$) bpos.

\medskip
A benchmark protocol is also considered.
This protocol is the same as the novel protocol
except for the relay-aided transmission mode.
Specifically, the relay-aided mode is the same as that
widely studied in the literature,
i.e., the source does not transmit at subcarrier $l$ during the second slot,
if subcarriers $k$ and $l$ are paired for the relay-aided transmission to user $u$.
In such a case, the maximum rate for the relay-aided transmission
over that subcarrier pair to user $u$ is equal to
$\Rate{\min\{\Gsrk\Pskone,\Gsuk\Pskone + \Grul\Prl\}}$ bpos.

\subsection{The sum rate maximization problem}

We assume there exists a central controller which knows precisely the CSI
$\{\Gsrk,\Gsuk,\Gruk|\forall\;k\}$. Before the data transmission,
the controller needs to find the optimum subcarrier assignment and power allocation
to maximize the sum rate of all users
for the adopted transmission protocol (which can be either the novel or benchmark protocol),
when the total power consumption is not higher than a prescribed value $\Ptot$.
Then, the controller can inform the source and the relay
the optimum RA to be adopted for data transmission.

It can be shown that the novel protocol leads to
a maximum sum rate greater than or equal to that for the benchmark one.
This can be proven as follows.
Note that the benchmark protocol is a special case of the novel one,
since it is equivalent to the novel one constrained with $\Psltwo=0$
if subcarrier $l$ is paired with a subcarrier for the relay-aided transmission.
Suppose the optimum subcarrier assignment
and power allocation have been found for the benchmark protocol.
By using the novel protocol with the same subcarrier assignment and power allocation,
the same sum rate can be achieved.
Obviously, the maximum sum rate for the novel protocol is greater than or equal to
that sum rate, namely the maximum sum rate for the benchmark protocol.

\section{RA Algorithm design}\label{sec:RA}

\subsection{Rate maximization for the relay-aided mode over a subcarrier pair}\label{sec:insight}

Assume subcarriers $k$ and $l$ are paired for the relay-aided mode
transmission to user $u$, and a sum power $P$ is used for this pair.
For the benchmark protocol,
it can be shown by using an intuitive method similar as
the one introduced in Appendix of \cite{WangTSP11} that
the maximum rate associated with the above optimum solution is equal to $\Rate{\oldGklu P}$ with
\begin{align}
\oldGklu = \left\{\begin{array}{ll}
               \frac{\Gsrk\Grul}{\diffGuk + \Grul}  & {\rm if\;}\min\{\Gsrk,\Grul\}> \Gsuk,  \\
               \min\{\Gsrk,\Gsuk\}                  & {\rm if\;}\min\{\Gsrk,\Grul\}\leq \Gsuk,
           \end{array}\right. \label{eq:oldGklu}
\end{align}
and the optimum $\Pskone$ is
\begin{align}
\Pskone = \left\{\begin{array}{ll}
               \frac{\Grul}{\diffGuk + \Grul}P   & {\rm if\;}\min\{\Gsrk,\Grul\}> \Gsuk,  \\
               P                                 & {\rm if\;}\min\{\Gsrk,\Grul\}\leq \Gsuk.
           \end{array}\right. \nonumber
\end{align}

To facilitate the derivation for the proposed protocol,
define $\diffGuk = \Gsrk - \Gsuk$ and $\sumGul = \Gsul + \Grul$.
To maximize the rate, the optimum $\Pskone$, $\Psltwo$ and $\Prl$ are the optimum solution for
\begin{align}
\max_{\Pskone,\Psltwo,\Prl}  &\hspace{0.2cm}  \min\{\Gsrk\Pskone, \SNRklu(\Pskone,\Psltwo,\Prl)\} \nonumber  \\
{\rm s.t.}                   &\hspace{0.2cm}  \Pskone+\Psltwo+\Prl = P,    \label{prob:novel-kl}\\
                             &\hspace{0.2cm}  \Pskone\geq0,\Psltwo\geq0,\Prl\geq0.      \nonumber
\end{align}


Using Cauchy-Schwartz inequality and an intuitive method 
similar as that introduced in Appendix of \cite{WangTSP11},
it can be shown that the optimum $\Pskone$, $\Psltwo$ and $\Prl$ 
for \eqref{prob:novel-kl} are
\begin{align}
\Pskone = \left\{\begin{array}{ll}
               \frac{\sumGul}{\diffGuk + \sumGul}P & {\rm if\;} \min\{\Gsrk,\sumGul\}> \Gsuk,  \\
               P                                 & {\rm if\;}   \min\{\Gsrk,\sumGul\}\leq \Gsuk,
           \end{array}\right. \nonumber
\end{align}
\begin{align}
\Psltwo = \left\{\begin{array}{ll}
               \frac{\Gsul}{\sumGul}\frac{\diffGuk}{(\diffGuk + \sumGul)}P & {\rm if\;} \min\{\Gsrk,\sumGul\}> \Gsuk,  \\
               0                                              & {\rm if\;} \min\{\Gsrk,\sumGul\}\leq \Gsuk,
           \end{array}\right. \nonumber
\end{align}
and $\Prl = P - \Pskone - \Psltwo$ (please see \cite{WangTSP13} for more details).
The maximum rate associated with the above optimum solution is equal to $\Rate{\newGklu P}$ with
\begin{align}
\newGklu = \left\{\begin{array}{ll}
               \frac{\Gsrk\sumGul}{\diffGuk + \sumGul}    & {\rm if\;}\min\{\Gsrk,\sumGul\}> \Gsuk,  \\
               \min\{\Gsrk, \Gsuk\}                       & {\rm if\;}\min\{\Gsrk,\sumGul\}\leq \Gsuk .
           \end{array}\right. \label{eq:newGklu}
\end{align}


\subsection{Formulation of the WSR maximization problem}

For the adopted protocol (which can be either the proposed or benchmark protocol),
we first define
\begin{align}
\Gklu = \left\{\begin{array}{ll}
                   \newGklu & {\rm if\;the\;proposed\;protocol\;is\;adopted},  \\
                   \oldGklu & {\rm if\;the\;benchmark\;protocol\;is\;adopted}.
           \end{array}\right. \nonumber
\end{align}

For any possible subcarrier assignment used by the adopted protocol,
suppose $m$ subcarrier pairs are assigned to the relay-aided transmission,
the unpaired subcarriers in the two slots can always be one-to-one associated
with each other to form $K-m$ virtual subcarrier pairs for the direct transmission.
Based on this observation, we define:
\begin{itemize}
\item
$\tklu\in \{0,1\}$ and $\Pklu\geq0$, $\forall\;k,l,u$.
$\tklu=1$ indicates that subcarrier $k$ is paired with subcarrier $l$
for the relay-aided transmission to user $u$.
When $\tklu=1$, $\Pklu$ is used as the total power
for the subcarrier pair $(k,l)$.

\item
$\tklab\in \{0,1\}$, $\pklab\geq 0$ and $\qklab\geq 0$, $\forall\;k,l,u$.
$\tklab=1$ indicates that subcarrier $k$ is assigned
in the direct transmission mode to user $a$
during the first slot, and so is subcarrier $l$ to user $b$ during the second slot.
When $\tklab=1$,  $\Pskone$ and $\Psltwo$ take the value of
$\pklab$ and $\qklab$, respectively.
\end{itemize}

Let us collect all indicator and power variables in the sets $\Iset$ and $\Pset$, respectively,
and define $\RA = \{\Iset,\Pset\}$. Every feasible RA scheme can be described by
an $\RA$ satisfying simultaneously
\begin{align}
&\tklu,\tklab\in \{0,1\}, \forall\;k,l,u,a,b,                              \label{eq:first-constr}\\
&\sum_{l}\left(\sum_{u}\tklu + \sum_{a,b}\tklab\right) = 1, \forall\;k,   \label{eq:second-constr}\\
&\sum_{k}\left(\sum_{u}\tklu + \sum_{a,b}\tklab\right) = 1, \forall\;l,   \label{eq:third-constr}\\
&\sum_{k,l,u,a,b} \left(\tklu\Pklu + \tklab(\pklab+\qklab)\right) \leq \Ptot, \label{eq:forth-constr}\\
&\Pklu\geq0, \pklab\geq0,\qklab\geq0,\forall\;k,l,u,a,b.        \label{eq:fifth-constr}
\end{align}

Given a feasible $\RA$, the maximum sum rate for the adopted protocol is
\begin{align}
f(\RA) = &\sum_{k,l,u,a,b}\big(\tklu \Rate{\Gklu \Pklu} + \\
         &\hspace{0.2cm}   \tklab\big(\Rate{\Gsak \pklab} + \Rate{\Gsbl\qklab}\big),   \nonumber
\end{align}
and the sum rate maximization problem is to solve
\begin{align}
\max_{\RA}  &\hspace{0.25cm} f(\RA)     \label{prob:firstRA}\\
{\rm s.t.}  &\hspace{0.25cm} \eqref{eq:first-constr}-\eqref{eq:fifth-constr} \nonumber
\end{align}
for a globally optimum $\RA$.
Obviously, \eqref{prob:firstRA} is a nonconvex mixed-integer nonlinear program.
To find a globally optimum $\RA$,
all indicator variables are first relaxed to be continuous within $[0,1]$, i.e.,
Then, we make the COV from $\Pset$ to $\newPset=\{\newPklu,\newpklab,\newqklab|\forall k,l,u,a,b\}$,
where every $\newPklu$, $\newpklab$ and $\newqklab$ satisfy, respectively,
\begin{align}
\newPklu = \tklu\Pklu, \newpklab=\tklab\pklab, \newqklab=\tklab\qklab.  \nonumber 
\end{align}

After collecting all variables into $\newRA = \{\Iset,\newPset\}$,
the RA problem can be rewritten as
\begin{align}
\max_{\newRA}  &\hspace{0.25cm} g(\newRA)     \label{prob:finalRA} \\
{\rm s.t.}     &\hspace{0.25cm} \tklu,\tklab\in [0,1], \forall\;k,l,u,a,b.    \label{eq:new-firstconstr}\\
               &\sum_{k,l,u,a,b} \left(\newPklu + \newpklab + \newqklab\right) \leq \Ptot, \label{eq:new-forthconstr}\\
               &\newPklu\geq0, \newpklab\geq0,\newqklab\geq0,\forall\;k,l,u,a,b,    \label{eq:new-fifthconstr}
\end{align}
where $g(\newRA)$ represents the maximum sum rate expressed as
\begin{align}
g(\newRA) =& \sum_{k,l,u,a,b} \big(\phi(\tklu,\newPklu,\Gklu)       \label{eq:new-WSR} \\
           &  + \phi(\tklab,\newpklab,\Gsak) + \phi(\tklab,\newqklab,\Gsbl)\big), \nonumber
\end{align}
and
\begin{align}
\phi(t,x,C) = \left\{\begin{array}{ll}
                     t \cdot \Rate{C \frac{x}{t}}   &   {\rm if\;}t>0,  \\
                     0                        &   {\rm if\;}t=0.
                   \end{array}\right.
\end{align}

Obviously, \eqref{prob:finalRA} is a relaxation of \eqref{prob:firstRA}.
We will find an (at least approximately) optimum solution for \eqref{prob:finalRA},
and show that the $\RA$ corresponding to this solution is still feasible,
and hence (at least approximately) optimum for \eqref{prob:firstRA}, 
which can be shown in a similar way as reported in \cite{WangTSP13}. 
To this end, note that that $\phi(t,p,G)$
is a continuous and concave function if $t\geq0$ and $x$,
because it is a perspective function of $\Rate{G p}$
which is concave of $p$ \cite{Convex-opt}.
As a result, $g(\newRA)$ is a concave function of $\newRA$
in its feasible domain for \eqref{prob:finalRA}.
This means that \eqref{prob:finalRA} is a convex optimization problem.
Apparently, it also satisfies the Slater constraint qualification,
therefore its duality gap is zero, which justifies the use of
the dual method to look for the globally optimum of \eqref{prob:finalRA}, 
denoted by $\newRA^\star$ hereafter.

To use the dual method, $\mu$ is introduced as a Lagrange multiplier
for the constraint \eqref{eq:new-forthconstr}.
The Lagrange relaxation problem (LRP) for \eqref{prob:finalRA} is
\begin{align}
\max_{\newRA}  &\hspace{0.25cm} L(\mu,\newRA) = g(\newRA) + \mu\bigg(\Ptot - \sumP(\newRA)\bigg)   \label{prob:LRP}\\
{\rm s.t.}     &\hspace{0.25cm} \eqref{eq:new-firstconstr},\eqref{eq:second-constr}-\eqref{eq:third-constr},
                       \eqref{eq:new-fifthconstr},  \nonumber
\end{align}
where $L(\mu,\newRA)$ is the Lagrangian of \eqref{prob:finalRA} and
$\sumP(\newRA)$ is the sum of all $\newPklu$, $\newpklab$ and $\newqklab$ in $\newRA$.
A global optimum of \eqref{prob:LRP} is denoted by $\newRA_\mu$.
The dual function is defined as $d(\mu)=L(\mu,\newRA_\mu)$.
In particular, $\Ptot-\sumP(\newRA_\mu)$ is a subgradient
of $d(\mu)$, i.e., it satisfies
$\forall\;\mu', d(\mu') \geq d(\mu) + (\mu'-\mu)(\Ptot-\sumP(\newRA_\mu))$,
and the dual problem is to find the dual optimum $\mu^\star = \arg\min_{\mu\geq0}d(\mu)$.

Since \eqref{prob:finalRA} has zero duality gap,
it satisfies two important properties.
First, $\mu^\star > 0$. This is because
$\mu^\star$ represents the sensitivity of the optimum objective value for \eqref{prob:finalRA}
with respect to $\Ptot$, i.e., $\frac{g(\newRA^\star)}{\Ptot} = \mu^\star$ \cite{Convex-opt}.
Obviously, $g(\newRA^\star)$ is strictly increasing of $\Ptot$, meaning that $\mu^\star>0$.
Second, $\mu$ and $\newRA$ are equal to $\mu = \mu^\star$ and $\newRA_\mu = \newRA^\star$,
if and only if $\newRA_\mu$ is feasible and $\mu(\Ptot-\sumP(\newRA_\mu)) = 0$ is satisfied
according to Proposition $5.1.5$ in \cite{Nonlinear-opt}.
Based on the above properties, the $\mu>0$ and $\newRA_{\mu}$
satisfying $\sumP(\newRA_{\mu})=\Ptot$ can be found as $\mu^\star$ and $\newRA^\star$.
Therefore, the key to developing a duality based algorithm consists of two procedures
to find $\mu^\star$ and $\newRA_\mu$, respectively.
We first introduce the one to find $\newRA_\mu$ as follows.

\subsubsection{Finding $\newRA_\mu$ when $\mu>0$}\label{sec:solve-LRP}

The following strategy is used to find $\newRA_\mu$ for \eqref{prob:LRP} when $\mu>0$.
First, the optimum $\newPset$ for \eqref{prob:LRP} with fixed $\Iset$ is found and
denoted by $\newPset_\Iset$.
Define $\newRA_\Iset = \{\Iset, \newPset_\Iset\}$.
Then we find the optimum $\Iset$ to maximizing $L(\mu,\newRA_\Iset)$
subject to \eqref{eq:new-firstconstr}, \eqref{eq:second-constr} and \eqref{eq:third-constr}.
Finally, $\newRA_\Iset$ corresponding to this optimum $\Iset$ can be taken as $\newRA_\mu$.

Suppose $\Iset$ is fixed, we find $\newPset_\Iset$ as follows.
Specifically, every $\newPklu$ in $\newPset_\Iset$ is equal to $0$
when $\tklu=0$. When $\tklu>0$, the optimum $\newPklu$ can be found by
using the KKT conditions related to $\newPklu$.
In summary, the optimum $\newPklu$ can be shown to be
$\newPklu = \tklu \Lambda(\mu,\Gklu)$,
where $\Lambda(\mu,G) = \left[\frac{\log_2{e}}{2\mu} - \frac{1}{G}\right]^+$.
In a similar way, the optimum $\newpklab$ and $\newqklab$
can be shown to be $\newpklab = \tklab\Lambda(\mu,G_{sa,k})$ and
$\newqklab = \tklab\Lambda(\wb,\mu,G_{sb,l})$.
respectively.
Using these formulas, $\newRA_\Iset = \{\Iset, \newPset_\Iset\}$
can be found. It can readily be shown that
\begin{align}
L(\mu,\newRA_\Iset) = \mu\Ptot + \sum_{k,l,u,a,b} \big(\tklu\Aklu + \tklab\Bklab\big)
\end{align}
where
\begin{align}
\Aklu =& \Rate{\Gklu\Lambda(\mu,\Gklu} - \mu\cdot\Lambda(\mu,\Gklu)   \nonumber
\end{align}
and
\begin{align}
\Bklab = &\Rate{\Gsak\Lambda(\mu,\Gsak)}- \mu\cdot\Lambda(\mu,\Gsak) +  \nonumber \\
         &\Rate{\Gsbl\Lambda(\mu,\Gsbl)}- \mu\cdot\Lambda(\mu,\Gsbl).    \nonumber
\end{align}

Finally, we find the optimum $\Iset$ for maximizing $L(\mu,\newRA_\Iset)$ subject to
\eqref{eq:new-firstconstr}, \eqref{eq:second-constr} and \eqref{eq:third-constr}.
This problem is equivalent to solving
\begin{align}
\max_{\Iset,\{\tkl|\forall\;k,l\}}
&\hspace{0.25cm} \sum_{k,l}\sum_{u,a,b} \big(\tklu\Aklu + \tklab\Bklab\big)   \nonumber\\
{\rm s.t.}    &\hspace{0.25cm} \sum_{l}\tkl = 1, \forall\;k,
                \hspace{0.1cm} \sum_{k}\tkl = 1, \forall\;l,    \label{prob:find-Ikl-first}\\
              &\hspace{0.25cm} \tkl = \sum_{u}\tklu + \sum_{a,b}\tklab,\forall\;k,l.       \nonumber\\
              &\hspace{0.25cm} \tklu\geq 0, \tklab\geq0,\forall\;k,l,u,a,b. \nonumber
\end{align}

Note that the inequality
\begin{align}
\sum_{u,a,b} \big(\tklu\Aklu + \tklab\Bklab\big) \leq \tkl \Ckl
\end{align}
holds where $\Ckl = \max\{\max_{u}\Aklu, \max_{a,b}\Bklab\}$.
Let us call $\Aklu$ as the metric for $\tklu$ and $\Bklab$ as the metric for $\tklab$.
This inequality is tightened when all entries of $\{\tklu,\tklab|\forall\;u,a,b\}$
are assigned to zero, except that the one with the metric equal to $\Ckl$ is assigned to $\tkl$.
Therefore, after the problem
\begin{align}
\max_{\{\tkl|\forall\;k,l\}}  &\hspace{0.25cm} \sum_{k,l}\sum_{u,a,b} \tkl\Ckl  \nonumber\\
{\rm s.t.}    &\hspace{0.25cm} \sum_{l}\tkl = 1, \forall\;k,  \hspace{0.1cm} \sum_{k}\tkl = 1, \forall\;l, \label{prob:find-Ikl-second}\\
              &\hspace{0.25cm} \tkl \geq 0, \forall\;k,l,   \nonumber
\end{align}
is solved for its optimum solution $\{\tkl^\star|\forall\;k,l\}$,
an optimum $\Iset$ for \eqref{prob:find-Ikl-first} can be constructed by assigning
for every combination of $k$ and $l$, all entries in $\{\tklu,\tklab|\forall\;u,a,b\}\subset\Iset$
to zero, except for the one with the metric equal to $\Ckl$ to $\tkl^\star$.

Most interestingly, \eqref{prob:find-Ikl-second} is a standard assignment problem,
hence $\{\tkl^\star|\forall\;k,l\}$ can be found efficiently by the Hungarian algorithm,
and every entry in $\{\tkl^\star|\forall\;k,l\}$ is either $0$ or $1$ \cite{Hungarian}.
After knowing $\{\tkl^\star|\forall\;k,l\}$, the optimum $\Iset$
can be constructed according to the way mentioned earlier.
Finally, the corresponding $\newRA_\Iset = \{\Iset,\newPset_\Iset\}$ is assigned to $\newRA_\mu$.
Note that the Hungarian algorithm to solve \eqref{prob:find-Ikl-second}
has a complexity of $O(K^3)$ \cite{Hungarian}, meaning that
the complexity of finding $\newRA_\mu$ is $O(K^3)$.

To find $\mu^\star$, an iterative method which updates $\mu$
with $\mu = [\mu - \delta(\Ptot-\sumP(\newRA_\mu))]^+$ can be used,
where $\delta>0$ is a prescribed step size \cite{Nonlinear-opt}.
However, this method converges very slowly,
since $\delta$ has to be very small to guarantee convergence.
To develop a faster algorithm, we first prove that
$\sumP(\newRA_\mu)$ is a decreasing function of $\mu>0$.
To this end, suppose $\mu_1\geq \mu_2 > 0$.
Since $\Ptot-\sumP(\newRA_{\mu_2})$ is a subgradient of $d(\mu)$ at
$\mu$, the inequalities
$d(\mu_1) \geq  d(\mu_2) + (\mu_1-\mu_2)(\Ptot-\sumP(\newRA_{\mu_2}))$
and
$d(\mu_2) \geq  d(\mu_1) + (\mu_2-\mu_1)(\Ptot-\sumP(\newRA_{\mu_1}))$
follow. As a result,
$(\mu_1 - \mu_2)(\Ptot-\sumP(\newRA_{\mu_1})) \geq  d(\mu_1) - d(\mu_2)   
\geq  (\mu_1-\mu_2)(\Ptot-\sumP(\newRA_{\mu_2}))$                   
holds, and thus $\sumP(\newRA_{\mu_1})) \leq \sumP(\newRA_{\mu_2})$,
meaning that $\sumP(\newRA_\mu)$ is a decreasing function of $\mu> 0$.
Based on the above property, a bisection method can be used to find the $\mu>0$
satisfying $\sumP(\newRA_\mu)=\Ptot$ as $\mu^\star$.

\begin{algorithm}
\caption{The algorithm to solve \eqref{prob:firstRA}.} \label{alg:dual}
\begin{algorithmic}[1]
\STATE  compute $\Gklu$, $\forall\;k,l,u$.
\STATE  $\mumin = 0$; $\mumax=1$; compute $\sumP(\newRA_{\mumax})$;
\WHILE{$\sumP(\newRA_{\mumax})\geq \Ptot$}
       \STATE  $\mumax = 2\mumax$; compute $\sumP(\newRA_{\mumax})$;
\ENDWHILE

\WHILE{$\mumax - \mumin > 0$}
       \STATE  $\mu = \frac{\mumax+\mumin}{2}$;  solve \eqref{prob:LRP} for $\newRA_\mu$;
       \IF{$ \Ptot-\epsilon\leq \sumP(\newRA_\mu)\leq \Ptot$}
           \STATE  go to line 12;
       \ELSIF{$\sumP(\newRA_\mu) > \Ptot$}
           \STATE  $\mumin = \mu$;
       \ELSE
           \STATE  $\mumax = \mu$;
       \ENDIF
\ENDWHILE
\STATE  compute the $\RA$ corresponding to $\newRA_{\mu}$ 
        and output it as an (at least approximately) optimum for (P1).
\end{algorithmic}
\end{algorithm}

The overall procedure to solve \eqref{prob:finalRA} 
is shown in Algorithm \ref{alg:dual}, 
where $\epsilon>0$ is a small prescribed tolerance.
It can be shown in a similar way as in \cite{WangTSP13} that 
the finally produced $\newRA_{\mu}$ is either equal to 
(if $\sumP(\newRA_\mu)=\Ptot$ is satisfied), or a close approximation 
(if $\Ptot-\epsilon\leq \sumP(\newRA_\mu)< \Ptot$ is satisfied)
for $\newRA^\star$. Moreover, the indicator variables in 
$\newRA_{\mu}$ are either $0$ or $1$. 
Therefore, the $\RA$ corresponding to $\newRA_{\mu}$
is either optimum or approximately optimum for \eqref{prob:firstRA}. 
After finding this optimum $\RA$, the optimum subcarrier assignment and source/relay
power allocation can computed accordingly. It can readily be shown that
Algorithm \ref{alg:dual} has a polynomial complexity with respect to $K$ and $U$.

\section{Numerical experiments}\label{sec:numexp}

\begin{figure}[h]
  \centering
     \includegraphics[width=3.5in]{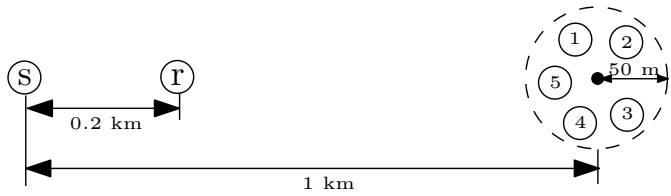}
  \caption{The relay-aided downlink OFDMA system considered in numerical experiments.}  \label{fig:simsys}
\end{figure}

Consider the relay-aided downlink OFDMA system illustrated in Fig. \ref{fig:simsys}.
$U=5$ users are served and the users are randomly and uniformly distributed
in a circular region of radius $50$ m.
To evaluate the maximum sum rate of the adopted protocol
for each fixed set of system parameters,
$500$ random realizations of channels are generated.
For each realization, the user coordinates are first randomly generated,
and then the channels are generated in the same way as in \cite{Wang11TSP-1}.

\begin{figure}[h]
  \centering
     \includegraphics[width=3.5in,height=2in]{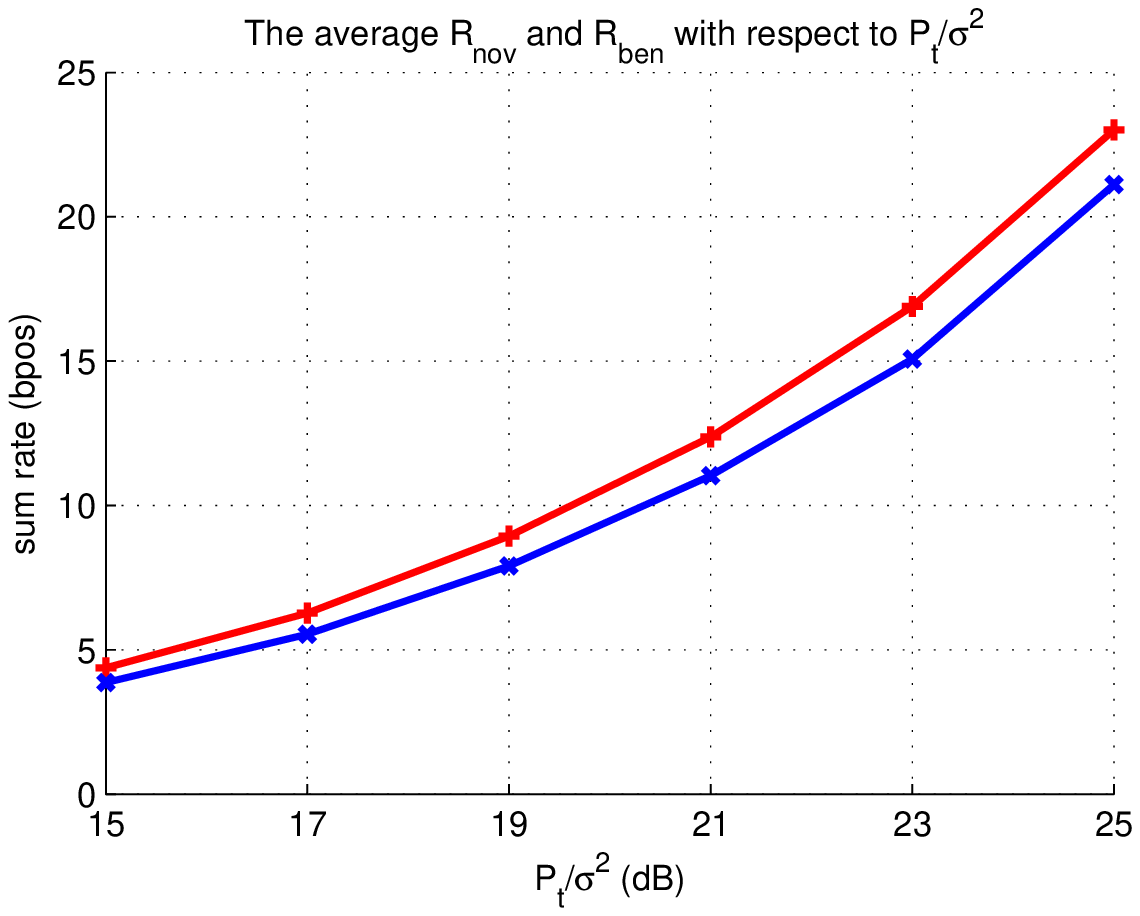}
  \caption{The average $\maxRnov$ and $\maxRben$
         when $K=32$ and $\Ptot/\sigma^2$ increases from $15$ to $25$ dB.}  \label{fig:results-Ptot}
\end{figure}

When $K=32$ and $\Ptot/\sigma^2$ increases from $15$ to $25$ dB,
the average $\maxRnov$ and $\maxRben$
which are the optimum sum rates for the two protocols, respectively,
are shown in Fig. \ref{fig:results-Ptot}.
It can be seen that the average $\maxRnov$ is always greater than the average $\maxRben$.
This illustrates the benefit of using the novel protocol.

\begin{figure}[h]
  \centering
  \includegraphics[width=3.5in,height=2in]{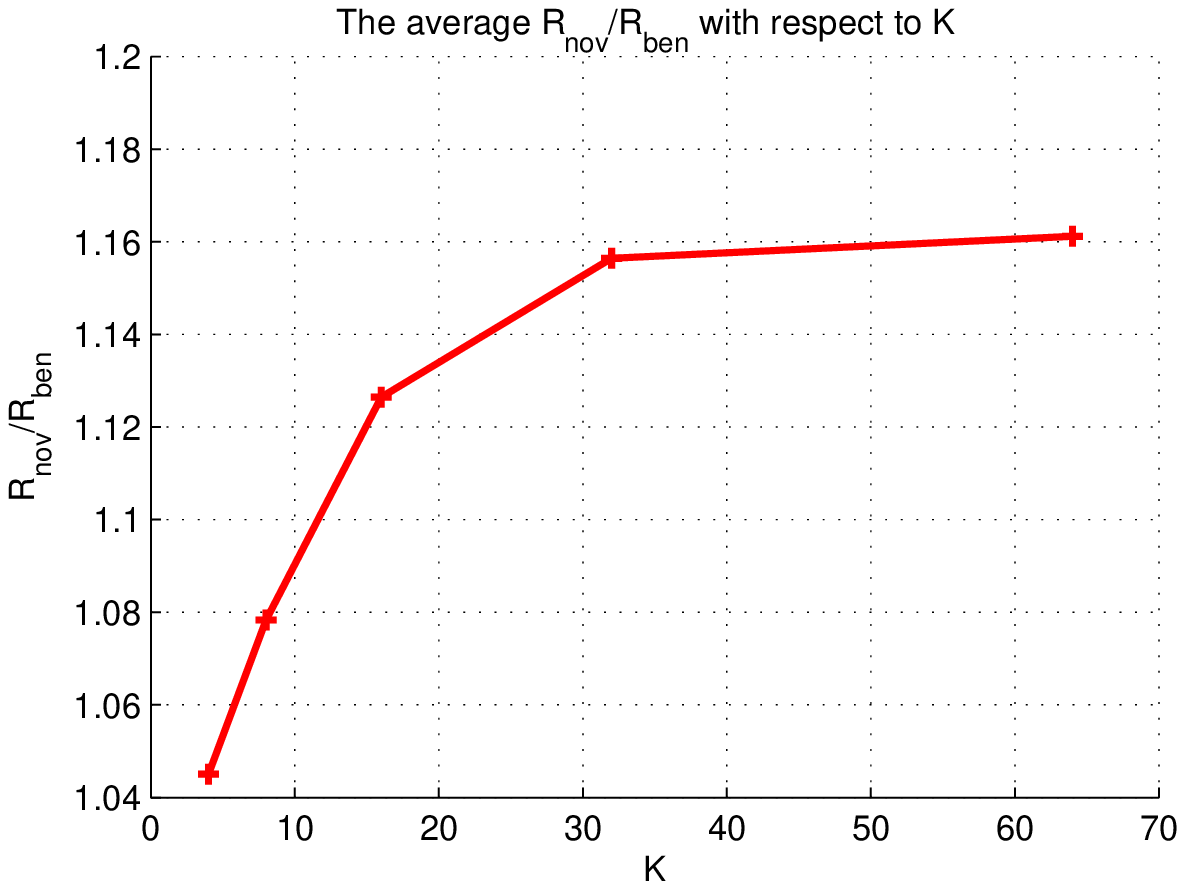}
  \caption{The average $\frac{\maxRnov}{\maxRben}$ when $\Ptot/\sigma^2 = 20$ dB
           and $K$ increases from $4$ to $64$.}  \label{fig:results-K}
\end{figure}

When $\Ptot/\sigma^2 = 20$ dB and $K$ increases from $4$ to $64$,
the average $\frac{\maxRnov}{\maxRben}$ are shown in Fig. \ref{fig:results-K}.
It can be seen that the average $\frac{\maxRnov}{\maxRben}$
is always greater than $1$,
and the benefit of using the novel protocol increases as $K$ increases.
This indicates that the proposed protocol can better
exploit the frequency-selective fading than the benchmark one
for optimizing the subcarrier assignment to maximize the sum rate,
especially when a big number of subcarriers is used.

\section{Conclusion}

We have proposed a novel subcarrier-pair based opportunistic DF relaying protocol
for downlink OFDMA transmission.
Note that the proposed protocol truly improves the DF relaying itself.
It is shown that the novel protocol leads to a maximum sum rate greater than
or equal to that for the benchmark one.
A polynomial-complexity RA algorithm has been developed for each protocol
to maximize the sum rate of all users.
Numerical experiments have illustrated that the novel protocol can
lead to a much greater sum rate than the benchmark one.


%



\ifCLASSOPTIONcaptionsoff
  \newpage
\fi



\bibliographystyle{IEEEtran}
\bibliography{subcarr-pairing}%



\end{document}